\documentclass[conference]{IEEEtran}
\IEEEoverridecommandlockouts

\usepackage{cite}
\usepackage{amsmath,amssymb,amsfonts}
\usepackage{algorithmic}
\usepackage{graphicx}
\usepackage{textcomp}
\usepackage{xcolor}
\usepackage{multirow}
\usepackage{float}
\usepackage{balance}
\usepackage{placeins}

\def\BibTeX{{\rm B\kern-.05em{\sc i\kern-.025em b}\kern-.08em
    T\kern-.1667em\lower.7ex\hbox{E}\kern-.125emX}}
\begin{document}

\title{IWN: Image Watermarking Based on Idempotency\\
}

\author{
\IEEEauthorblockN{Kaixin Deng}
}


\maketitle

\begin{abstract}
In the expanding field of digital media, maintaining the strength and integrity of watermarking technology is becoming increasingly challenging. This paper, inspired by the Idempotent Generative Network (IGN), explores the prospects of introducing idempotency into image watermark processing and proposes an innovative neural network model—the Idempotent Watermarking Network (IWN). The proposed model, which focuses on enhancing the recovery quality of color image watermarks, leverages idempotency to ensure superior image reversibility. This feature ensures that, even if color image watermarks are attacked or damaged, they can be effectively projected and mapped back to their original state. Therefore, the extracted watermarks have unquestionably increased quality. The IWN model achieves a balance between embedding capacity and robustness, alleviating to some extent the inherent contradiction between these two factors in traditional watermarking techniques and steganography methods.
\end{abstract}


\section{Introduction}
The growing popularity of digital multimedia has made image watermarking a key research area for information hiding and copyright protection\cite{cox2007digital}. Image watermarking embeds covert data into digital media, providing a secure method for upholding copyright and data integrity. This allows individuals and organizations to assert ownership and authenticity, preventing unauthorized copying or tampering. However, existing watermark extraction methods often suffer quality degradation when the watermarked image is attacked or damaged. This paper focuses on improving the resilience and recovery quality of color image watermarks under such adverse conditions.

The Idempotent Generative Network (IGN)\cite{shocher2023idempotent} is a novel architecture based on the mathematical property of idempotency, where repeated application of an operation leaves the result unchanged. Formally, for a function $f$, idempotency means $f(f(...f(x)...))=f(x)$. In IGN, this ensures the network produces consistent outputs even when reprocessing its own output, providing stability and consistency. For image processing, IGN can generate images with specific features that remain stable through multiple network passes. The potential applications of idempotency are significant, ensuring both information consistency and processing stability. Furthermore, IGN's idempotency follows the principles of orthogonal projection - an idempotent linear operator that preserves specific data components while eliminating others\cite{strang2012linear}. This allows IGN to retain essential input features while filtering out extraneous details, enabling efficient compression and representation without compromising content integrity.

To the best of our knowledge, this work pioneers the use of idempotency and projection mechanisms for stable image watermarking. We propose the Idempotent Watermarking Network (IWN), which inherits IGN's core idempotency. For input $x$ (an image with or without a watermark), IWN's output remains unchanged after repeated projections and mappings:
\begin{equation}
IWN(IWN(...IWN(x)...))=IWN(x)
\label{eq:my_equation1}
\end{equation}

IWN assumes the original image $x$, watermark $w$, and watermarked image $x_w$ share the same distribution. Idempotency ensures watermark stability - for an unwatermarked image, erroneous watermark introduction is prevented; for a watermarked image $x_w$, the watermark persists through projection-mapping and can be restored even if attacked or damaged. This effectively maps damaged watermarked images back to their original watermarked state. Beyond ensuring watermark stability, this significantly improves IWN's reversibility when handling attacked or damaged watermarks, enhancing the reliability and utility of watermarking for information hiding. 

The IWN algorithm proposed in this paper achieves a breakthrough in this regard, skillfully balancing embedding capacity and robustness. It not only embeds a watermark image of the same size as the original but also demonstrates exceptional robustness. 

\section{Related Work}

\subsection{Generative Networks}

Generative Adversarial Networks\cite{goodfellow2014generative} (GANs) is one kind of deep learning framework comprising two competing neural networks: the Generator and the Discriminator. The Generator works to simulate the distribution of real data in order to generate new, realistic samples. In contrast, the Discriminator's task is to distinguish between real data and data generated by the Generator. During the training process, these two networks promote each other and go hand in hand. The efficacy and stability of GANs depend on their structural design\cite{karras2019style,pan2023drag}. 

Diffusion models\cite{sohl2015deep} have garnered widespread attention in the field of generative models, particularly demonstrating outstanding performance in the generation of images, audio, and text\cite{song2020denoising,yang2023diffusion,ho2022classifier}. These models draw inspiration from the diffusion process within the realm of non-equilibrium thermodynamics. In the forward process, the model iteratively transforms data into a Gaussian distribution via a predefined Markov chain. This conversion is achieved by incrementally introducing noise to the data over numerous steps. Conversely, the reverse process involves denoising, starting with noise and progressively removing it to reconstruct the data.

The Idempotent Generative Network (IGN), with its distinctive strengths, successfully addresses the limitations inherent in current approaches. IGN is designed to circumvent the multi-step inference process required by autoregressive methods. It is able to generate robust outputs in a single step, similar to single-step inference models like GANs. The fundamental principle underlying the operation of IGN lies in the mapping of the source distribution (such as Gaussian noise) to the target distribution (such as real images). This is achieved by ensuring that instances of the target distribution map to themselves, while instances of the source distribution map onto a predefined target manifold. This strategy, which involves mapping the source distribution to the target distribution while ensuring idempotency, enables the model to complete the generation of outputs in a single step, maintaining a consistent latent space.

\subsection{Techniques for Watermark Processing}

The domain of the embedded data determines the classification of watermarking techniques\cite{cox2002digital,hartung1999multimedia,begum2020digital}. These techniques are primarily divided into two categories: spatial domain\cite{su2018robust} watermarking and transform domain watermarking. The spatial domain watermarking technique embeds watermark information by minutely adjusting the pixel values of images\cite{lai2010digital}, offering benefits such as lower computational complexity and increased processing efficiency. However, this method's drawback lies in the watermark's relative lack of stability and resistance to interference. Transform domain watermarking techniques, which embed watermark information by applying specific signal transformations to images, significantly enhance the robustness and anti-attack capabilities of the watermark. 

While both image watermarking and steganography algorithms involve embedding information in images, their objectives and methods differ. Image watermarking primarily serves copyright protection and authentication purposes, emphasizing robustness. Notably, many image watermarking algorithms, especially spatial domain methods like least significant bit modification, typically require the embedded watermark to be much smaller than the original image to maintain imperceptibility. These methods often alter only a small portion of pixels or bit planes, minimizing changes to the original image while ensuring the watermark's presence. In contrast, steganography focuses on covert communication, prioritizing concealment and embedding capacity over robustness. Advanced image steganography algorithms can embed larger amounts of secret information, sometimes approaching the size of the original image. This high-capacity embedding capability gives steganography an advantage in scenarios requiring transmission of substantial covert data. However, these high-capacity steganographic methods often struggle to extract high-quality secret information when faced with various attacks, reflecting their lack of robustness.

\section{Methods}

\subsection{Method of Watermark Embedding} 

This paper discusses the implementation of the IWN model, which embeds watermarks using the Discrete Cosine Transform\cite{ahmed1974discrete} (DCT) technique. DCT is an efficient linear transformation method that converts signals from the time domain to the frequency domain, and it is widely used in signal compression and filtering. The core of the DCT is its capacity to break down signals into a linear combination of discrete cosine base functions. By modulating the coefficients of these base functions for filtering, it accomplishes noise reduction and data storage optimization. In DCT, the primary energy and key information of an image are concentrated in the low-frequency area, while the details of the image are mainly distributed in the mid-to-high-frequency area. In the context of image compression, these visually less impactful mid-high frequency details are often compressed, and these areas are also the parts most susceptible to damage during the compression process. 

The watermark embedding process begins with a two-dimensional Discrete Cosine Transform (DCT) on the original image $x$ and the watermark image $w$. The DCT transformations of the original image and the watermark image are denoted as $DCT(x)$ and $DCT(w)$ respectively. The watermarked DCT coefficients are obtained by adding the weighted DCT coefficients of the watermark to the DCT coefficients of the original image. This embedding process can be mathematically expressed as follows:
\begin{equation}
DCT(x_w) = DCT(x) + \alpha \times DCT(w)
\label{eq:my_equation2}
\end{equation}
Here, $\alpha$ denotes the watermark embedding strength, aiming to balance the watermark's visibility and robustness. The watermark extraction process involves performing a two-dimensional DCT transformation on the watermarked image $x_w$ and the original image $x$. The watermark information is extracted by subtracting the DCT coefficients of the original image from those of the watermarked image. The extracted watermark DCT coefficients, denoted by $DCT(w_{extracted})$, are restored to the watermark image $w_{extracted}$ through the inverse DCT transformation (IDCT). This extraction and restoration process is represented by the following mathematical expressions:
\begin{equation}
DCT(w_{extracted}) = \frac{DCT(x_w) - DCT(x)}{\alpha}
\label{eq:my_equation3}
\end{equation}
\begin{equation}
w_{extracted} = F^{-1}(DCT(w_{extracted}))
\label{eq:my_equation4}
\end{equation}

The two-dimensional DCT transform $F(u, v)$, representing the frequency domain representation of the image, is mathematically defined as:
\begin{equation}
\begin{aligned}
A &= \cos\left[\frac{\pi(2x + 1)u}{2N}\right] \cos\left[\frac{\pi(2y + 1)v}{2M}\right] , \\
F(u, v) &= \frac{2}{\sqrt{NM}} C(u) C(v) \sum_{x=0}^{N-1} \sum_{y=0}^{M-1} f(x, y) \cdot A
\end{aligned}
\label{eq:my_equation5}
\end{equation}
Here, $C(u)$ represents the normalization coefficients, defined as:
\begin{equation}
C(u) = \begin{cases}\frac{1}{\sqrt{2}} & \text{if } u = 0\\1 & \text{otherwise}\end{cases}
\label{eq:my_equation6}
\end{equation}

The mathematical expression for the inverse DCT transform (IDCT), which restores the watermark image from its DCT coefficients, is defined as:
\begin{equation}
\begin{aligned}
A &= \cos\left[\frac{\pi(2x + 1)u}{2N}\right] \cos\left[\frac{\pi(2y + 1)v}{2M}\right] ,\\
f(x, y) &= \frac{2}{\sqrt{NM}} \sum_{u=0}^{N-1} \sum_{v=0}^{M-1} C(u) C(v) F(u, v) \cdot A
\end{aligned}
\label{eq:my_equation7}
\end{equation}
In this formulation, $N \times M$ represents the size of the image block, $(x, y)$ are the spatial coordinates of the image, and $(u, v)$ are the coordinates in the transformation domain. The function $f(x, y)$ denotes the pixel value at the spatial coordinates $(x, y)$, while $F(u, v)$ denotes the DCT coefficients at the frequency domain coordinates $(u, v)$.

\subsection{Backbone Network}

\begin{figure*}[!htbp]
    \centering
    \includegraphics[width=1\linewidth]{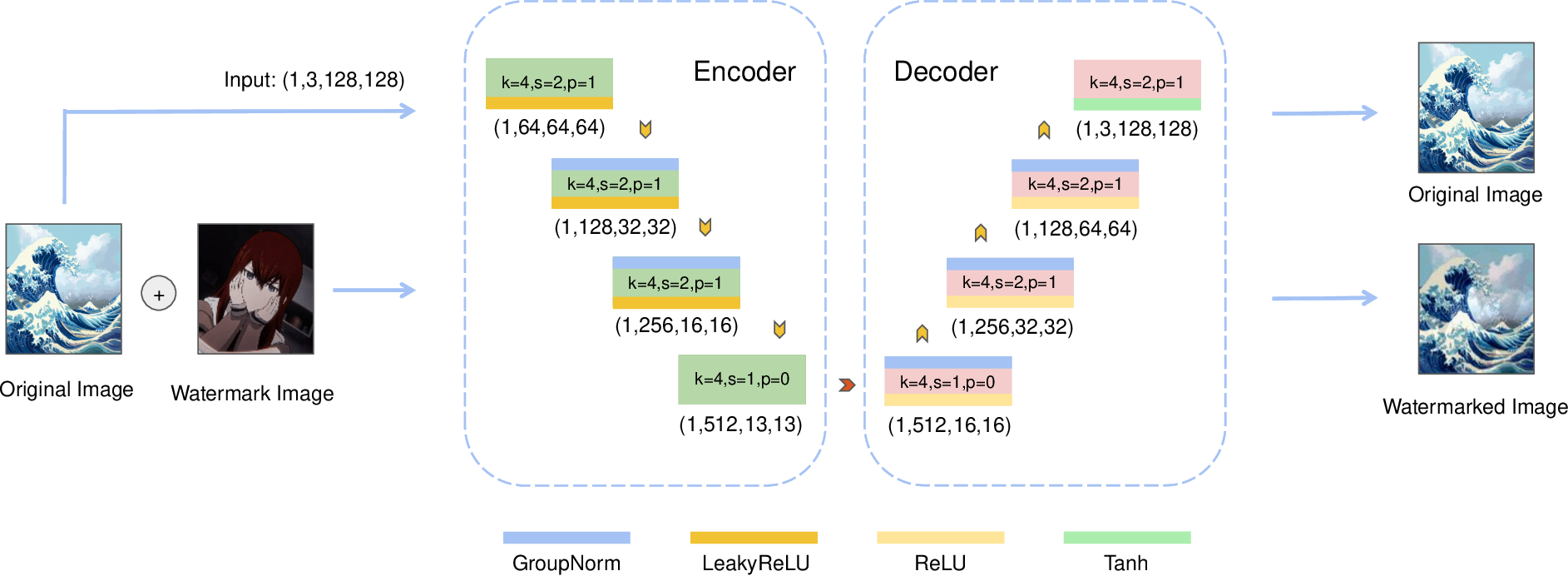}
    \caption{\textbf{The structure of the backbone network.} Both the input and output images of the network are of size (128, 128).}
    \label{fig:backbone}
\end{figure*}

The IWN model employs a widely used encoder-decoder structure, such as the U-Net structure often used in medical image segmentation\cite{deng2023improved}, as shown in Fig. \ref{fig:backbone}. Upon completion of training, the model ensures that the original image retains its watermark-free state, even following multiple iterations of model mapping. Simultaneously, it maintains the integrity of the watermark information within the watermarked images during the same mapping process. The encoder part gradually reduces the spatial dimensions and increases the feature dimensions through convolutional layers and activation functions, thereby extracting high-level features of the image. GroupNorm\cite{wu2018group} is utilized to optimize training stability. The decoder part gradually restores the spatial dimensions of the image and reduces the feature dimensions through transposed convolutional layers, ultimately outputting the image through the Tanh activation function. This design strategy gives the network excellent performance in both feature extraction and image reconstruction.

\subsection{Optimization Objective}

The goal of the IWN model is to ensure that the state of the image watermark after projection mapping by the model is highly consistent with the target distribution. This goal is achieved through the following five loss functions.

\subsubsection{Reconstruction Loss} 
The IWN model is based on the important assumption that the noise distribution, denoted by $P_z$, and the target distribution, denoted by $P_x$, occupy the same space. This is achieved through the use of reconstruction loss, which ensures that data samples extracted from the target distribution (denoted by $x_i$) are mapped back to their original distribution (denoted by $P_x$). This setup rationalizes the application of the mapping function $f$ to a particular sample $x$ extracted from the target distribution $P_x$. The goal of the mapping function $f$ is to maintain the stability of the instances on the target manifold. This can be achieved by applying the mapping function $f$ to map the samples $x$ from the target distribution $P_x$ to itself. Mathematically, this is expressed as $f(x) = x$. This maintains the state of the original image, $x$, and the watermarked image, $x_w$, unchanged, and accomplishes the reversibility of the original image. This process defines the predicted manifold as the set of all instances that are mapped to themselves or to their neighbors by the function $f$. This process can be represented by the following equation:
\begin{equation}
L_{rec} = \mathop{\min}\limits_{\theta} \sum_i (\left| x_i - f_\theta(x)_i \right| + \left| x_i - f_\theta(x_w)_i \right|)
\label{eq:my_equation8}
\end{equation}
Here, $\theta$ represents the parameters of the network $f_\theta$, $x$ is the original image, and $x_w$ is the image after watermarking.

\subsubsection{Idempotency Loss}
The second objective of the IWN model is to achieve the idempotency of the model, as expressed by the mathematical equation $f(f(...f(z)...))=f(z)$. This implies that the output after multiple consecutive applications of the model is consistent with the output after a single application, thereby enhancing the model's stability and reliability. This characteristic enables the model to effectively map instances from different distributions $P_z$ to the estimated manifold. Each instance $z$ is mapped by the function $f$ to its corresponding position on the manifold. Where instance $z$ is a severely damaged image embedded with a watermark. When optimizing $f(f(z))-f(z)=0$, two distinct gradient path flows exist, both of which contribute during the optimization process. The optimization of $f(z)$ in the inner layer of $f(f(z))$ is desired, as this allows for a better mapping of $f(z)$ to the target distribution. However, when optimizing the $f(z)$ in the outer layer of $f(f(z))$, it leads to the expansion of the target distribution, which is an unintended result. In extreme scenarios, the trained network $f$ for all inputs $z$ results in $f(z)=z$, signifying that the input is noise, and the output remains as noise. This approach allows for a more accurate mapping of $f(z)$ to the target distribution. However, when optimizing the $f(z)$ in the outer layer of $f(f(z))$, it can lead to the expansion of the target distribution, which is an unintended result. In extreme cases, the trained network $f$ for all inputs $z$ results in $f(z)=z$, indicating that the input is noise and the output remains as noise. Consequently, it is desirable to update the parameters of the inner layer $f$ while maintaining the parameters of the outer layer $f$. The outer layer $f$ may be regarded as a frozen copy of the current state of $f$. This objective effectively recovers damaged watermark images by idempotency loss, thereby enhancing the robustness of hidden information recovery. The mathematical expression is:
\begin{equation}
L_{idem} = \mathop{\min}\limits_{\theta} \sum_i \left| f_\theta(z)_i - f{_{\theta'}}(f_\theta(z))_i \right|
\label{eq:my_equation9}
\end{equation}
In this context, $f_{\theta'}$ represents a copy of $f_{\theta}$, where $\theta'$ is the parameter of $f_{\theta'}$. It should be noted that while $\theta'$ and $\theta$ are equivalent in value, they are distinct entities. This copy is employed to ensure that the gradients of the outer network $f_{\theta'}$ do not update the parameters of the inner network $f_{\theta}$.

\subsubsection{Compactness Loss} 
Although the first two objectives ensure that $x$ and $f(z)$ are positioned on the target manifold, they do not constrain other possible elements on that manifold. Therefore, the third objective, compactness loss, in the IWN model aims to minimize the subset of instances that map onto themselves. This ensures both the precision of the mapping and the compactness of the target manifold. One can address the issue of target distribution expansion by freezing the parameters of the inner layer $f$, and only updating the parameters of the outer layer $f$. This can be achieved by applying the mathematical expression:
\begin{equation}
L_{tight} = -\mathop{\min}\limits_{\theta} \sum_i \left| f_{\theta'}(z)_i - f{_\theta}(f_{\theta'}(z))_i \right|
\label{eq:my_equation10}
\end{equation}
It is notable that $L_{tight}$ may result in significant alterations in instances that are already of high quality, which could potentially compromise the stability of the model. Furthermore, this loss function encourages an increase in the distance between the model's output and input, which could lead to excessive gradients and model instability. To address these issues, the IWN model employs a hyperbolic tangent function based on the current reconstruction loss to smoothly constrain the value of $L_{tight}$:

\begin{equation}
\begin{aligned}
L_{\text{tight}} &= \tanh\left(\frac{\min_{\theta} \sum_i |g(z)_i|}{-rL_{\text{rec}}}\right) \cdot L_{\text{rec}} 
\end{aligned}
\label{eq:my_equation11}
\end{equation}
Where $g(z)_i = f_{\theta'}(z)_i - f_{\theta}(f_{\theta'}(z))_i$ and $r$ is a constant, determining the degree of constraint imposed by $L_{tight}$.

\subsubsection{Watermark Image Loss}

The fourth objective of the IWN model is to ensure that the watermark image, following projection mapping by the model, can accurately extract the watermark, maintaining the stability and integrity of the watermark. It is desired that the watermark extracted after being processed by $f(x_w)$ closely resembles the watermark extracted directly from $x_w$. To this end, we introduce watermark image loss, which focuses on preserving watermark information. The mathematical expression for this is as follows:

\begin{equation}
\begin{aligned}
L_{\text{watermark}} &= \min_{\theta} \sum \left| \mathcal{T}\{E_{\text{watermark}}\} \right| , \\
E_{\text{watermark}}[m,n] &= e(x_w)[m,n] - e(f(x_w))[m,n]
\end{aligned}
\label{eq:watermark_loss}
\end{equation}
Where $e(x_w)$ and $e(f(x_w))$ represent the results of the original watermark image $x_w$ and the processed watermark image $f(x_w)$ after the watermark extraction operation $e(\cdot)$, and $\mathcal{T}\{\cdot\}$ denotes the 2D transform:
\begin{equation}
\mathcal{T}\{E_{\text{watermark}}\} = \sum_{m=0}^{M-1} \sum_{n=0}^{N-1} E_{\text{wm}}[m,n] \cdot e^{-2\pi i (\frac{mk}{M} + \frac{nl}{N})}
\label{eq:fourier_transform}
\end{equation}

Here, $M$ and $N$ are the dimensions of the image in terms of rows and columns. $m$ and $n$ are the indices of the image in the rows and columns, respectively. The coordinates $k$ and $l$ represent the frequency domain. This formula quantifies the difference in the frequency domain between the original watermark image and the watermark image processed by the model. By minimizing this difference, the parameters $\theta$ of the model f are optimized, ensuring the stability and integrity of the watermark are preserved during the processing.
\subsubsection{Original Image Loss}
The fifth objective of the IWN model requires that the original image, after being processed by the model, does not introduce watermarks that would compromise the fidelity and purity of the original image. Therefore, we propose the original image loss, with the mathematical expression as follows:
\begin{equation}
\begin{aligned}
L_{origin} &= \mathop{\min}\limits_{\theta} \sum \left| \mathcal{T}\{E_{\text{origin}}\} \right| , \\
E_{origin}[m,n] &= \left( e(x)[m, n] - e(f(x))[m, n] \right)
\end{aligned}
\label{eq:my_equation13}
\end{equation}
In this expression, $e(x)$ and $e(f(x)$ represent the results of the original image $x$ and the processed image $f(x)$ after a specific extraction operation $e(\cdot)$, with the meanings of the other variables consistent with those in the watermark image loss. $\mathcal{T}\{\cdot\}$ denotes the 2D transform:
\begin{equation}
\mathcal{T}\{E_{\text{origin}}\} = \sum_{m=0}^{M-1} \sum_{n=0}^{N-1} E_{o}[m,n] \cdot e^{-2\pi i \left(\frac{mk}{M} + \frac{nl}{N}\right)}
\label{eq:fourier_transform}
\end{equation}

Combining the aforementioned optimization components, the final loss function is formulated as follows:
\begin{equation}
L = L_{rec} + {\lambda_i} L_{idem} + {\lambda_t} L_{tight} + {\lambda_w} L_{wm} + {\lambda_o} L_{o}
\label{eq:my_equation14}
\end{equation}
In this context, the symbol $\lambda$ represents the weights corresponding to each component of the loss function. This comprehensive optimization strategy allows the IWN model to maintain the stability of the intrinsic image properties. The use of idempotency allows the effective recovery of watermarks from color images that have been attacked or damaged. As a result, the quality of the extracted watermark is significantly improved. However, there may be potential conflicts or trade-offs between different objectives due to the variety of optimization objectives. The selection of loss weights is crucial to ensure that the model can effectively balance the various optimization objectives and achieve the desired performance. In subsequent work, adaptive weight adjustment strategies such as dynamic weight scaling can be considered to help the model automatically adjust the loss weights.

\section{Experiments}



\subsection{Training Parameters}

For the IWN model, the image size was set to (128,128) and the batch size to 1. The optimization algorithm utilized was Adam, with an initial learning rate set to 1e-4. The model underwent a total of 10 training epochs. In terms of loss function weights, the weights for reconstruction loss, idempotency loss, and compactness loss were set to 25, 20, and 2.5, respectively. The constant r in the compactness loss function was set to 1.5. Furthermore, the weights for both the watermark image frequency domain loss and the original image frequency domain loss were 0.1. It is notable that, in order to ensure comprehensive training, each loaded original and watermark image was duplicated 1200 times and used as training data.

\subsection{Evaluation Metrics}

In this study, we primarily utilized the PSNR and SSIM\cite{hore2010image,wang2004image} metrics to assess the performance of the model. PSNR is a metric commonly employed to assess the quality of image reconstruction. It is frequently utilized in image compression and various image processing domains. PSNR is primarily based on the concept of MSE (Mean Squared Error), which represents the ratio of the maximum possible power of a signal to the power of the noise affecting it. The mathematical expression for PSNR is as follows:
\begin{equation}
\text{PSNR} = 10 \cdot \log_{10} \left( \frac{\text{MAX}^2}{\text{MSE}} \right)
\label{eq:my_equation15}
\end{equation}
Where, $\text{MAX}$ is the maximum possible value of image pixels. For instance, for an 8-bit image, $MAX = 255$. MSE (Mean Squared Error) is the mean squared error between the original image and the compressed image, with the following mathematical expression:
\begin{equation}
\text{MSE} = \frac{1}{mn} \sum_{i=0}^{m-1} \sum_{j=0}^{n-1} [I(i, j) - K(i, j)]^2
\label{eq:my_equation16}
\end{equation}
Where, $I(i, j)$ is the pixel value of the original image at position $(i, j)$, and $K(i, j)$ is the pixel value of the compressed image at the same position, $m$ and $n$ are the height and width of the image, respectively. The higher the PSNR value, the closer the quality of the reconstructed image is to the original image.

The SSIM (Structural Similarity Index) metric is a measure of the similarity between two images. Unlike the PSNR (Peak Signal-to-Noise Ratio) metric, SSIM considers the structural information of the images. Its objective is to provide an evaluation that is closer to the quality perceived by human vision. The mathematical expression for SSIM is as follows:
\begin{equation}
\text{SSIM}(x, y) = \frac{(2\mu_x \mu_y + c_1)(2\sigma_{xy} + c_2)}{(\mu_x^2 + \mu_y^2 + c_1)(\sigma_x^2 + \sigma_y^2 + c_2)}
\label{eq:my_equation17}
\end{equation}
The two images to be compared are designated by $x$ and $y$. The mean values of $x$ and $y$, respectively denoted by $\mu_x$ and $\mu_y$, are the arithmetic means of the respective variables. The variances of $x$ and $y$, respectively denoted by $\sigma_x^2$ and $\sigma_y^2$, are the square of the standard deviations of the respective variables. The covariance of $x$, denoted by $\sigma_{xy}$, is the correlation coefficient between $x$ and $y$. Furthermore, the values of $c_1 = (k_1L)^2$ and $c_2 = (k_2L)^2$ are employed as small constants to maintain stability. The value of $L$ represents the dynamic range of pixel values, while the default values of $k_1$ and $k_2$ are utilized. When the SSIM value is equal to one, it signifies that the two images exhibit complete structural identity.

\begin{figure}[htbp]
    \centering
    \includegraphics[width=1.0\linewidth]{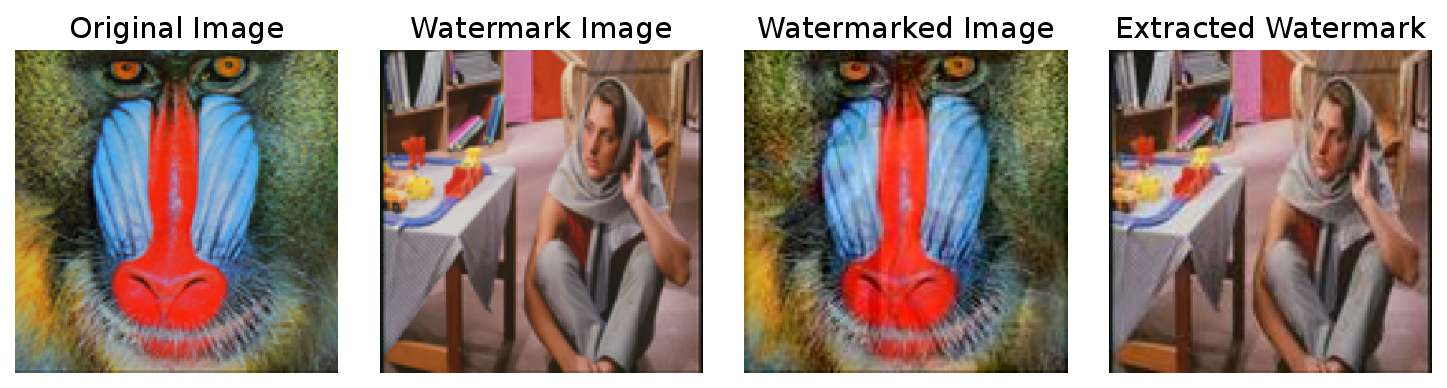}
    \caption{\textbf{A demonstration of the DCT algorithm.}}
    \label{fig:dct_res}
\end{figure}
\subsection{Experimental Results}

\subsubsection{DCT Watermark Embedding and Extraction}
We begin by demonstrating the efficacy of the DCT algorithm in the context of watermark embedding and extraction. The results indicate that the original image, watermark image, image after watermark embedding, and the effect of watermark extraction are shown in Fig. \ref{fig:dct_res}.

The results demonstrate that although the DCT algorithm is capable of implementing basic watermark embedding and extraction operations, certain limitations remain. Consequently, future work will adopt more advanced and complex algorithms to enhance the concealment and anti-interference ability of the watermark.

\subsubsection{Validation of Model Optimization Objectives}
\begin{figure}[htbp]
    \centering
    \includegraphics[width=1\linewidth]{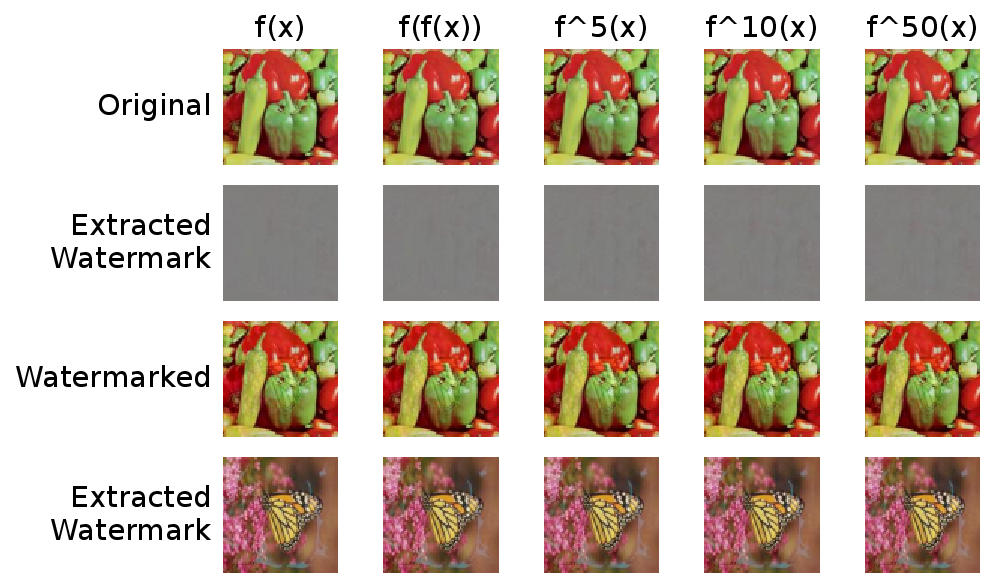}
    \caption{\textbf{Results of multiple projection mapping using the model.} Even after processing 50 times using the IWN, the extracted watermarked image does not change much.}
    \label{fig:idempotency_iwn}
\end{figure}

Subsequently, the core optimization objectives of the IWN model were validated. These objectives were to ensure that the unwatermarked image $x$ maintained its unwatermarked state after multiple projections and mappings through the model, and that the watermarked image $x_w$ consistently retained its watermark state after multiple projections and mappings through the model. The visualization results are shown in Fig. \ref{fig:idempotency_iwn}.

The results demonstrate that after 50 projection mappings, the original image does not erroneously introduce watermarks. The image with the embedded watermark consistently exhibits stable watermark extraction quality even after 50 projection mappings. Furthermore, both the original image and the image with the embedded watermark are able to maintain their own state stability, which aligns with our optimization objectives.

\begin{figure}[htbp]
    \centering
    \includegraphics[width=1\linewidth]{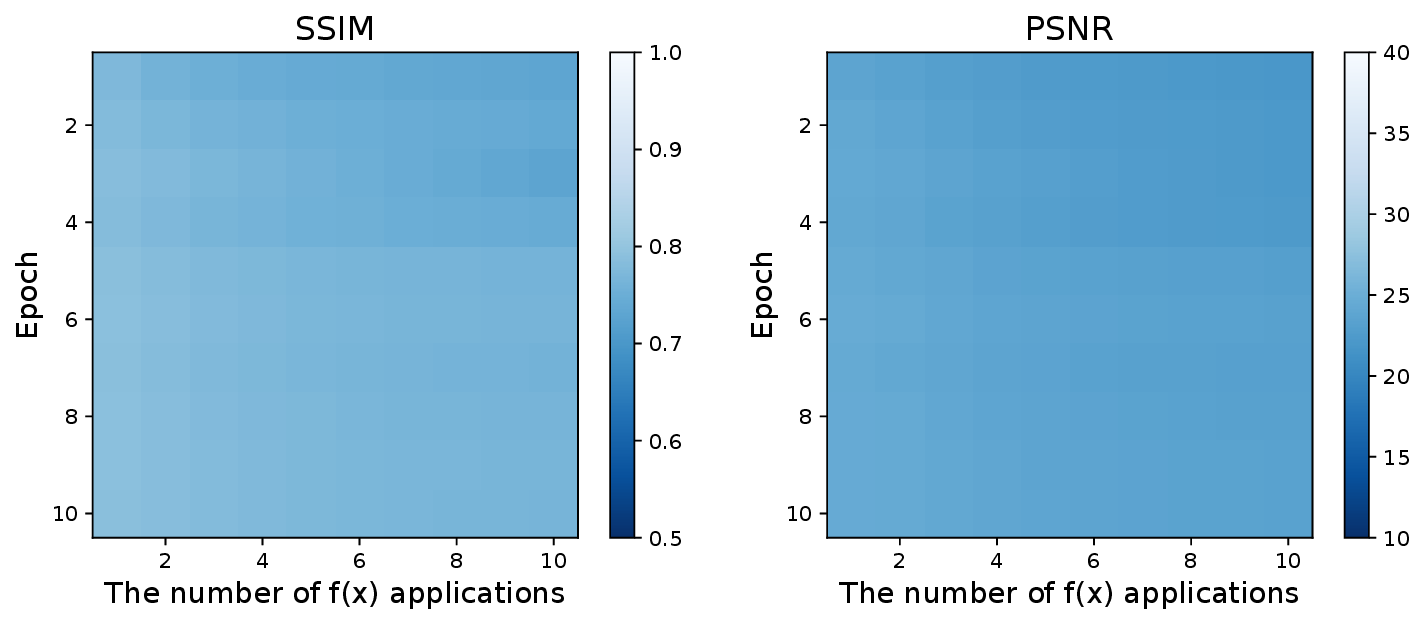}
    \caption{\textbf{Heat map of the impact of mapping operations on image quality using the IWN model projection during the training phase.}}
    \label{fig:heatmap}
\end{figure}

We conducted further analysis on the Set14\cite{zeyde2012single} dataset to evaluate the impact of the projection mapping operation of the IWN model on the quality of the embedded watermarked images at different training stages. The evaluation was performed by calculating the PSNR and SSIM metrics against the original watermarked image and presenting the results in the form of a heatmap, as shown in Fig. \ref{fig:heatmap}.

The results demonstrate that during the initial stages of training, there is a decline in image quality with an increasing number of projection mappings. However, as the training progresses, the image quality remains stable, maintaining inherent stability regardless of the number of projection mapping operations applied. This effectively achieves the desired result of $f(f(...f(z)...))=f(z)$.

\subsubsection{Recovery Ability for Damaged Watermark Images}
Furthermore, we investigated the efficacy of the IWN model in processing color image watermarks that have been subjected to attack or damage. Five distinct attack methodologies were employed, as detailed in Table. \ref{tab:attack-methods}.
\begin{table}[htbp]
\centering
\caption{Attack Methods and Parameters}
\label{tab:attack-methods}
\begin{tabular}{|c|c|c|}
\hline
\textbf{ID} & \textbf{Attack Method} & \textbf{Parameters} \\
\hline
A & Mosaic & Random 50\% region \\
B & Gaussian Noise & Noise std: 0.5; Mean: 0 \\
C & Salt and Pepper Noise & SP ratio: 0.5; Noise amount: 0.15 \\
D & JPEG Compression & Quality value: 1 (max: 100)\\
E & Gaussian Filter & Kernel size: 3; Sigma: 1.0; 3 times \\
\hline
\end{tabular}
\end{table}

\begin{figure}[htbp]
    \centering
    \includegraphics[width=1\linewidth]{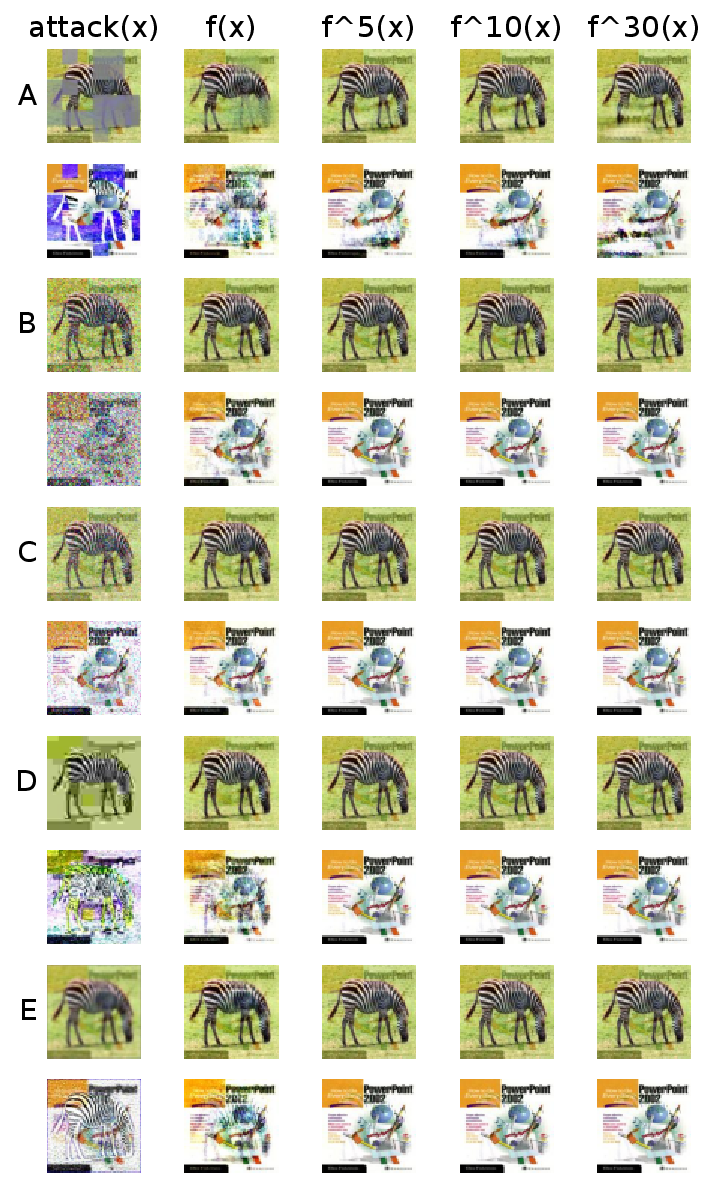}
    \caption{\textbf{Results of processing the attacked watermarked image using the IWN model.} Here, zebra in the Set14 dataset is the original image to which we embedded the watermarked image ppt3.}
    \label{fig:attacked_result}
\end{figure}

We have carried out the projection mapping process of the attacked watermarked image with 1, 5, 10 and 30 times IWN model and the results are shown in Fig. \ref{fig:attacked_result}

\begin{figure}[htbp]
    \centering
    \includegraphics[width=1\linewidth]{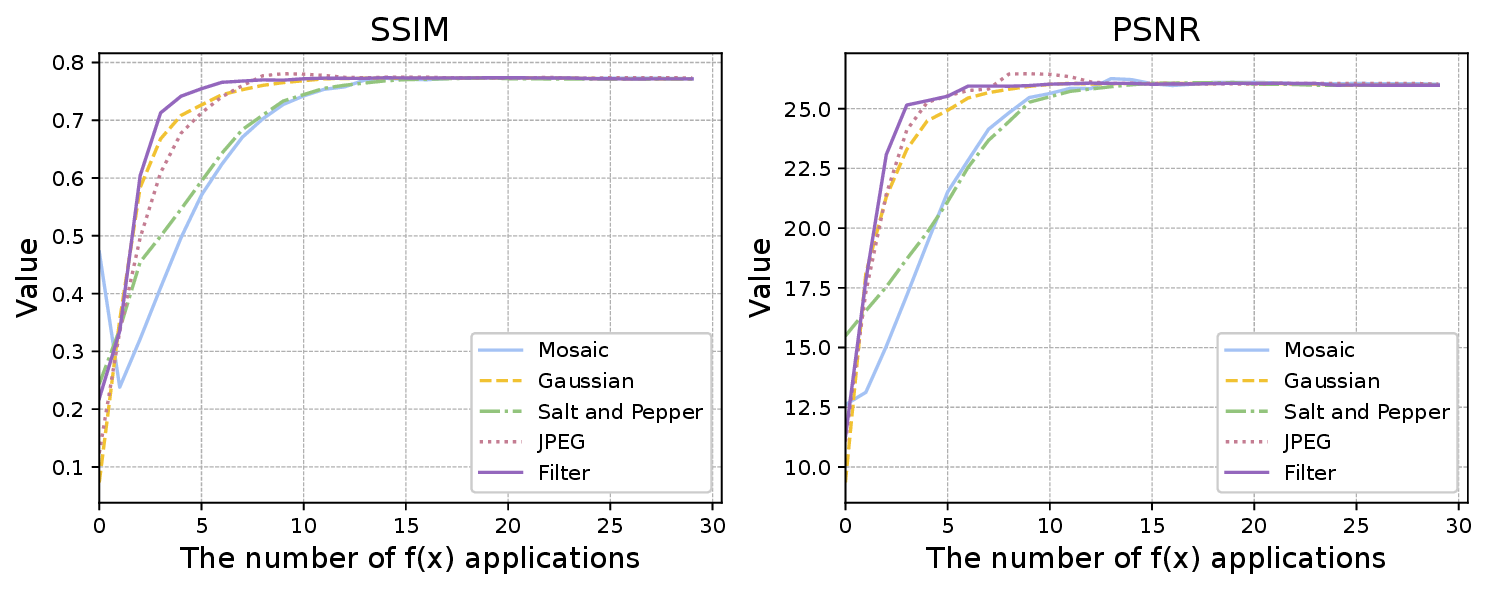}
    \caption{\textbf{Results after multiple projection processing using the IWN model for different attack scenarios.}}
    \label{fig:ssim_psnr}
\end{figure}

To investigate the influence of the number of projection mappings performed by the IWN model on image quality, a visual analysis was conducted on the SSIM and PSNR metrics calculated against the original watermark images following projection processing using the IWN model under five attack scenarios. The results are presented in Fig. \ref{fig:ssim_psnr}.

The experimental results demonstrate that the IWN model demonstrates a robust ability to restore attacked or damaged color image watermarks to their near-original states, thereby considerably enhancing the quality of the extracted watermarks to levels akin to those of the original watermark images. Additionally, under specific conditions, the IWN model achieves substantial improvements in the quality of damaged watermark images, attaining an SSIM index close to 0.8 and a PSNR index close to 30. Despite these advances, the IWN model currently encounters difficulties in achieving an optimal mapping from source distribution instances to the target manifold through a single projection mapping, due to the complexity of its optimization objectives. 

\subsubsection{Comparison Experiments}
This paper presents comparative experiments between the IWN model and state-of-the-art image watermarking and steganography algorithms in the Set14 dataset. Experimental results reveal that the IWN model successfully embeds a watermark image of equal size to the original while maintaining exceptional robustness. The IWN model demonstrates a clever balance between embedding capacity and robustness. Notably, some algorithms \cite{kasa2023vanGonography} require the embedded watermark information to be significantly smaller than the original image, so these algorithms cannot embed a watermark image of the same size as the original. The results are shown in Table \ref{tab:algorithm-comparison}.

\begin{table*}[htbp]
\centering
\caption{Comparison of Different Algorithms under Various Attack Types}
\label{tab:algorithm-comparison}
\resizebox{\textwidth}{!}{%
\begin{tabular}{|l|cc|cc|cc|cc|cc|}
\hline
\multirow{2}{*}{\textbf{Algorithm}} & \multicolumn{2}{c|}{\textbf{Mosaic}} & \multicolumn{2}{c|}{\textbf{Gaussian Noise}} & \multicolumn{2}{c|}{\textbf{Salt \& Pepper}} & \multicolumn{2}{c|}{\textbf{JPEG}} & \multicolumn{2}{c|}{\textbf{Gaussian Filter}} \\
\cline{2-11}
 & \textbf{SSIM} & \textbf{PSNR} & \textbf{SSIM} & \textbf{PSNR} & \textbf{SSIM} & \textbf{PSNR} & \textbf{SSIM} & \textbf{PSNR} & \textbf{SSIM} & \textbf{PSNR} \\
\hline
Steganography \cite{kelvins2023steganography}& 0.495 & 11.95 & 0.013 & 9.31 & 0.456 & 15.76 & 0.036 & 7.84 & 0.010 & 9.31 \\
HiNet \cite{Jing_2021_ICCV}& 0.804 & 19.90 & 0.017 & 6.00 & 0.038 & 6.61 & 0.107 & 11.64 & 0.066 & 11.80 \\
PUSNet \cite{Li_2024_CVPR}& 0.662 & 14.12 & 0.025 & 7.29 & 0.034 & 7.49 & 0.018 & 7.05 & 0.024 & 7.48 \\
Ours & 0.809 & 22.38 & 0.866 & 24.48 & 0.831 & 23.78 & 0.671 & 19.66 & 0.873 & 24.53 \\
\hline
\end{tabular}%
}
\end{table*}

\subsubsection{Ablation Studies}
This study conducted ablation experiments on the proposed structure, as shown in Table \ref{tab:network-structure-comparison}. By comparing three network structures of different scales, we found that the original version achieved a good balance between model complexity and performance. It employs a structure with 4 encoder layers and 4 decoder layers, totaling 9.58M parameters. The original version reached 22.97 dB in PSNR and 0.810 in SSIM, showing significant improvement over the small version. Although the large version demonstrated superior performance metrics, its 38.95M parameters lead to substantial computational resource consumption and extended training time. Taking all factors into account, the original version may be the optimal choice for practical applications, particularly in scenarios requiring a balance among performance, computational resources, and model generalization capacity.

\begin{table*}[htbp]
\centering
\caption{Comparison of Network Structures for Different Model Versions}
\label{tab:network-structure-comparison}
\resizebox{\textwidth}{!}{%
\begin{tabular}{|l|c|c|c|c|c|c|}
\hline
\multirow{2}{*}{\textbf{Setting}} & \multicolumn{2}{c|}{\textbf{small version}} & \multicolumn{2}{c|}{\textbf{origin version}} & \multicolumn{2}{c|}{\textbf{large version}} \\
\cline{2-7}
 & \textbf{Encoder} & \textbf{Decoder} & \textbf{Encoder} & \textbf{Decoder} & \textbf{Encoder} & \textbf{Decoder} \\
\hline
Number of layers & 3 & 3 & 4 & 4 & 5 & 5 \\
\hline
Channel configurations & 64, 128, 256 & 256, 128, 3 & 64, 128, 256, 512 & 512, 256, 128, 3 & 64, 128, 256, 512, 1024 & 1024, 512, 256, 128, 3 \\
\hline
Kernel sizes & 4, 4, 4 & 4, 4, 4 & 4, 4, 4, 4 & 4, 4, 4, 4 & 4, 4, 4, 4, 4 & 4, 4, 4, 4, 4 \\
\hline
Stride configurations & 2, 2, 1 & 1, 2, 2 & 2, 2, 2, 1 & 1, 2, 2, 2 & 2, 2, 2, 2, 1 & 1, 2, 2, 2, 2 \\
\hline
Padding configurations & 1, 1, 0 & 0, 1, 1 & 1, 1, 1, 0 & 0, 1, 1, 1 & 1, 1, 1, 1, 0 & 0, 1, 1, 1, 1 \\
\hline
Batch norm configurations & F, T, F & T, T, F & F, T, T, F & T, T, T, F & F, T, T, T, F & T, T, T, T, F \\
\hline
Total parameters & \multicolumn{2}{c|}{2.24M} & \multicolumn{2}{c|}{9.58M} & \multicolumn{2}{c|}{38.95M} \\
\hline
PSNR & \multicolumn{2}{c|}{7.56} & \multicolumn{2}{c|}{22.97} & \multicolumn{2}{c|}{28.40} \\
\hline
SSIM & \multicolumn{2}{c|}{0.066} & \multicolumn{2}{c|}{0.810} & \multicolumn{2}{c|}{0.897} \\
\hline
\end{tabular}%
}
\end{table*}

\section{Conclusion And Prospects}

This paper introduces the concept of idempotency to watermark image processing by proposing the Idempotent Watermarking Network (IWN). The IWN model incorporates the idempotency of the Idempotent Generative Network (IGN), ensuring that non-watermarked images remain unaltered after multiple processing rounds. In contrast, watermarked images can be restored to their original state following damage or attack. The IWN exhibits remarkable resilience and fidelity, particularly in the face of common attacks. Its robust recovery capabilities are evident. While the IWN model proposed in this paper has achieved preliminary results, advancing and deepening image watermarking technology is an ongoing process. Future research may be conducted in the following areas:
\begin{itemize}
    \item Currently, the IWN model is unable to completely map a damaged watermarked image back to its original state through a single projection operation. Future research could explore the possibility of mapping instances to the target manifold through a single projection.

    \item The watermark processing method employed in this study is based on the discrete cosine transform (DCT). Subsequent improvements could be made by employing more advanced algorithms with lower perceptibility and stronger resistance to attacks.

    \item In light of the growing demand for processing speed in practical applications, future research could focus on optimizing the computational efficiency of the model, reducing processing time, and achieving faster model training and a more lightweight or efficient model architecture.

\end{itemize}

\FloatBarrier
\bibliographystyle{IEEEtran}
\bibliography{IEEEabrv,citations}


\end{document}